\documentclass[10pt,conference,twocolumn]{IEEEtran}
\usepackage{eucal}
\usepackage{graphicx}
\usepackage{times}
\usepackage{latexsym}
\usepackage{amsmath}
\usepackage{amssymb}
\usepackage{epsfig}
\usepackage{cite}
\usepackage{cases}
\usepackage{amsfonts}
\usepackage{indentfirst}
\usepackage{epsfig}
\usepackage{amsopn}
\usepackage{bm}
\usepackage{algorithmic}
\usepackage{algorithm}
\usepackage{subfig}
\usepackage{eufrak}

\IEEEoverridecommandlockouts

\begin{document}

\title{Efficient Image Transmission Through Analog Error Correction\thanks{Supported by National Science Foundation under Grants No. CMMI-0928092, CCF-0829888 and OCI-1122027. This paper appears in 2011 IEEE Multimedia Signal Processing (MMSP) Workshop, Oct. 2011.}}

\author{Yang Liu, \ \ \ Jing Li (Tiffany) \ \ and \ \ Kai Xie \\
 Electrical and Computer Engineering Department, \\
 Lehigh University, Bethlehem, PA 18015\\
 Email: \{yal210, jingli, kax205\}@ece.lehigh.edu\vspace*{-0.6cm}
}

\maketitle



\begin{abstract}
This paper presents a new paradigm for image transmission through analog error correction codes. Conventional schemes rely on digitizing images through quantization (which inevitably causes significant bandwidth expansion) and transmitting binary bit-streams through digital error correction codes (which do not automatically differentiate the different levels of significance among the bits). To strike a better overall performance in terms of  transmission efficiency and quality, we propose to use a single analog error correction code  in lieu of digital quantization, digital code and digital modulation. 
The key is to get analog coding right. We show that this can be achieved by cleverly exploiting an elegant ``butterfly'' property of chaotic systems. Specifically, we demonstrate a tail-biting triple-branch baker's map code and its maximum-likelihood decoding algorithm. Simulations show that the proposed analog code can actually outperform digital turbo code, one of the best codes known to date! The results and findings discussed in this paper speak volume for the promising potential of analog codes, in spite of their rather short history. 
 

\end{abstract}


\section{Introduction}

\PARstart{W}{e} live in a digitized information world, but many
of the physical sources, such as sound and images, are by nature
analog. To transmit analog sources using digital communication
systems require the signals to first of all be A/D
(analog-to-digital) converted, which involves sampling and
quantization. Sampling, the operation that transforms a signal
from  ``continuous in time'' to ``discrete in time'', is
reversible, i.e. the original continuous-time signal can be
reconstructed from the discrete samples loss-free, provided
the samples were taken at (or above) the Nyquist rate. However,
quantization, the operation that transforms a signal from
``continuous in amplitude'' to ``discrete in amplitude'', is
irreversible, i.e., the distortion caused by rounding off the
signal amplitude cannot be recovered after quantization.
To keep down the granularity error in general requires a large
number of quantization levels and/or high-dimension (vector) quantization. Since vector quantizers are very 
challenging to design, and usually require the knowledge of
high-order source statistics (e.g. joint probability distribution of the $n$th order for $n$-dimension quantization) 
which may not be easily available, real-world systems 
tend to use simpler scalar quantizers with many 
levels, at the cost of a large bandwidth expansion.

However, quantization error and bandwidth expansion are not all
the problems. Another practical issue concerns the labeling. 
An $n$-level quantization scheme takes
$\lceil log_2 n\rceil$ bits to label each level. Regardless of
what labeling scheme is used (Gray, natural or
mixed label),  different bits in the label will have different
levels of importance, but this natural hierarchy is not
reflected in the communication channel which treats all the bits
equal. For example, consider a 256-level monochrome image via natural labeling, where each pixel takes 8 bits to represent. An error in the least significant bit causes the pixel to be distorted by only one gray level, whereas an error in the most significant bit causes a drastic distortion of 128 levels!   
To avoid (wasteful) over-protection of some bits and/or (disastrous) under-protection of others, one must employ unequal error protection (UEP), 
but to provision the right protection to bits in accordance with their individual importance is nevertheless challenging. 
Many design issues arise, such as how much more important and hence how much more protection one bit deserves in comparison to another, 
what error correction codes (lengths, rates, error correction capabilities) are appropriate for each, and how to
balance the rates between quantization, error correction   
and modulation. Most of these issues are difficult to quantify
or optimize.

\begin{figure}[htb] 
\vspace{-0.2in}
\centerline{
\includegraphics[width=2.8in]{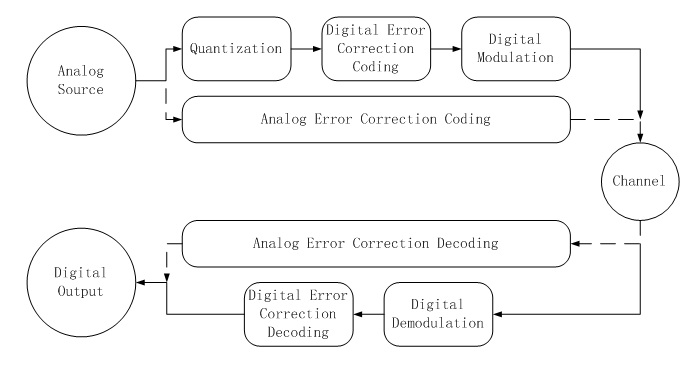}
\vspace{-0.2in}
}
\caption{A single analog error correction code in lieu of the combination of quantization, digital error correction code and digital modulation.}
\label{fig:analogDigital}
\end{figure}

While all of the above issues appear like a fact-of-life that one
has to accept, they actually need not be so. They stand in the way
only because we force analog signals into a {\it digitalized}
transmission paradigm. Consider an analog alternative that leaves
out quantization altogether and directly transmits discrete-time
continuous-valued analog signals (see Fig.
\ref{fig:analogDigital}), then all the quantization problems 
would be gone \cite{bib:analogTransmit}. The only obstacle, however, is that efficient analog error correction codes (AECC) are hard to find.

One exciting result we wish to report here is that practical and
efficient AECC {\it can} be designed and analog transmission can
be made reliable {\it and} simple! A much under-studied topic
(especially compared to the prolific literature on digital error
correction codes),
the notion of {\it analog error correction codes}, or, {\it real
number codes}, actually dates back to the early 80's, when 
Marshall and Wolf independently introduced the concept \cite{bib:Mars81}\cite{bib:Wolf83a}.
 Early ideas of analog codes were a natural
outgrowth of digital codes, by extending conventional digital
codes from the finite field to the real-valued or complex-valued
field (i.e. symbols from a very large finite field can approximate
real values). This has resulted in, for example, discrete Fourier
transform (DFT) codes (a subclass of which become analog BCH codes
and analog Reed-Solomon codes) \cite{bib:Mars81}\cite{bib:Wolf83a}\cite{bib:Wolf83b}, 
discrete cosine transform (DCT) codes \cite{bib:DCT}, and graph-based analog codes \cite{bib:vardy}. Although
linear codes dominate the short literature of analog codes like
they do in digital codes, linear analog codes are not nearly as
powerful as linear digital codes\footnote{A 
performance lower bound (in terms of mean square error) was
recently established for an arbitrary $(n,k)$ linear analog code, and 
it is shown that a carefully-designed {\it nonlinear} analog code
can easily beat this bound 
\cite{bib:linearAnalog}.}, and {\it nonlinear} analog codes are
true cause for excitement \cite{bib:linearAnalog}.

Nonlinear analog codes rely on nonlinear transforms to encode
analog data. Of particular interest is {\it chaotic analog codes} 
(CAC), a special class of nonlinear AECC that make essential use
of chaotic systems to transform signals. Chaotic systems are
nonlinear dynamical systems with bounded state spaces exhibiting a
topological mixing feature \cite{bib:chaos}. They are widely existent in the
natural world and the engineering world (e.g. climate change,
mechanical vibration, acoustic signals and  ecology systems are all chaotic systems), and many of them can be realized or emulated
using simple electric circuits (e.g. chua's circuit \cite{bib:chua}). 
Despite the rich variety of formalities, chaotic systems share an important common
property of {\it high sensitivity to the initial state}. Popularly
dubbed the ``butterfly effect,'' this property states that a small
perturbation to the initial state of a chaotic system will cause a 
huge difference later on. Although this butterfly effect is in
general viewed as a system penalty, it can actually be cleverly
exploited to satisfy the {\it distance expansion} property required by a
good error correction code. Specifically, if one treats the
initial state of a chaotic system as the source (to be encoded),
and treats some later states as the codeword (having been
encoded), then the chaotic system naturally enacts an error
correction encoder that successfully magnifies the small
differences among the source sequences (i.e. distance expansion).

This elegant feature was first exploited by Chen and Wornell in
the late nineties, when they proposed the first-ever chaotic
analog code, the {\it tent map code} \cite{bib:tentmapcode}. 
Using a single {\it tent map} (a simple one-dimension
chaotic function) as the encoder, they demonstrated the
feasibility of constructing error correction codes using chaotic systems. Sadly,
however, their code did not perform nearly as well as 
digital codes, and hence the wonderful idea exposed therein slept
for a decade before it was recently picked up by Xie, Tan, Ng
and Li \cite{bib:cat}. Leveraging the successful experience from
digital turbo codes, i.e. building long, powerful codes by
concatenating shorter, weaker codes, \cite{bib:cat} succeeded in
constructing {\it chaotic analog turbo} (CAT) codes by parallelly 
concatenating two tent maps. Just like turbo codes significantly outperform convolutional codes, CAT codes significantly outperform tent map codes.
 The work of \cite{bib:mirror} further extends the parallel concatenation idea to 2-dimension chaotic maps and proposed {\it mirrored baker's map codes}. 

In this paper we present a further generalization of the
idea of constructing a long, powerful system using a set of shorter, weaker
components. The key is to carefully leverage the strength of one another to cover up their individual weaknesses. Specifically, we
develop a class of {\it tail-biting analog codes} (TAC) based on 
2-dimensional chaotic maps. Previous work has considered a two-branch mirrored construction \cite{bib:mirror}. Here we present a constructive example that
engages three branches of {\it baker's maps} in a looped
tail-biting manner, and discuss its maximum-likelihood (ML) decoding
algorithm. 
To support our proposal of analog image transmission, we apply our analog
codes in image transmission, and compare it with  
the state-of-the-art digital systems (i.e. turbo codes). Simulations reveal
 a surprisingly good performance achieved by our analog codes, which, for practical purposes, is considerably better than digital turbo codes! 
The result is particularly exciting, considering that 1) the proposed analog coding system incurs considerably less complexity, memory and delay than the turbo coding system, and 2) turbo codes, the well-known class of capacity-approaching codes, represent the culmination of 70 years of mature digital coding research, whereas the research of analog codes is still at a very early stage. 

\section{Principles and System Model}
\label{sec:system}

Error correction is based on a simple but profound idea of {\it
distance expansion}. Through a linear or nonlinear map (i.e.
encoding), the {\it source space} in which elements have
relatively small separations and are therefore easily distorted,  
is transformed to a {\it code space} of a larger 
dimension, where elements have (much) 
larger separation and can therefore tolerate (much) larger 
perturbation. 
 Distance expansion is generally achieved by adding redundancy and hence incurs 
bandwidth expansion.  
An $(n,k)$ code that encodes a length-$k$ source sequence to a
length-$n$ codeword has increased the bandwidth consumption from
$k$ units to $n$ units. The code rate, defined as
$r\!=\!k/n\!<\!1$, provides a measure of the amount of bandwidth
expansion.

Chaotic systems are described by nonlinear chaotic functions whose
Lyapunov exponents $>1$. A discrete-time chaotic function
describes the time evolution of the state vector $\mathbf{z}$, 
\begin{align}
\mathbf{z}[i] =F(\mathbf{z}[i-1]), 
\end{align}
where $\mathbf{z}[0]$ denotes the initial state (seed). A rate $1/n$
code can be realized by feeding source symbols to the chaotic
function as the seed $\mathbf{z}[0]$, and collecting $(n\!-\!1)$
consecutive states.

The proposed tail-biting analog codes are based on 2-dimension chaotic maps, whose state vector $\mathbf{z}$ has a dimension of 2. Previous chaotic analog codes, such as the tent map code proposed in \cite{bib:tentmapcode} and the chaotic analog turbo code proposed \cite{bib:cat}, are based on  1-dimension chaotic maps (e.g. the tent map), and hence have an source block size of only 1, that is, the resultant code is always an $(n,1)$ code which encodes one source symbol to a codeword of $n$ encoded symbols. From the information theory, a larger block size in general provides a richer context, a better ``diversity'' and hence a better performance. However, high-order-dimension discrete-time chaotic functions, those that have relatively simple structures and hence allow for practical detection with manageable complexity, are very difficult to find. For this reason, we resort to 2-dimension chaotic maps, and employ a looped tail-biting structure to connect them. That is, we can take $k$ branches of 2-dimension chaotic map, take a block of $k$ source symbols, $u_1, u_2, ..., u_k$, and feed the source symbols as initial states to the $k$ branches:  \vspace{-0.6cm}
\begin{align}
\{u_1,u_2\} & \mbox{\ for Branch 1},\nonumber \\
\{u_2,u_3\} & \mbox{\ for Branch 2},\nonumber \\
\cdots & \cdots \nonumber \\
\{u_{k-1},u_k\} & \mbox{\ for Branch\ }k\!-\!1,\nonumber \\
\{u_{k},u_1\} & \mbox{\ for Branch\ }k.\nonumber 
\end{align}

\vspace{-0.2cm}
Below we detail an 
example of triple branches. 

\section{Triple-Branch Tail-Biting Baker's Map Codes}
\label{sec:decoding}

In what follows,  
we will use regular fonts to denote scalars (e.g. $x_1$), and bold
fonts to denote vectors and matrices (e.g. $\mathbf{x_1}$).
$\mathbf{x_1}{}_m^l$ denotes the vector $(x_1[m], x_1[m\!+\!1],
\cdots, x_1[l])$, and $\mathbf{x_1}$ is short for
$\mathbf{x_1}{}_0^{n-1}$.

\subsection{Encoder}

We consider using folded baker's map,  a simple, 2-dimension 
chaotic map from a unit square to itself, as the base function.
The baker's map is named after a kneading operation that
bakers apply to dough: the dough is cut in half, and one half is
folded over and stacked on the other, and compressed. It is
nonlinear but piece-wise linear, and presents a 2-dimension 
analogy of the tent map:
\begin{align}
&\ \{ x[i],y[i]\} \nonumber\\
 =& F(\,\{x[i\!-\!1],y[i\!-\!1]\}\,) \nonumber \\
=&\left\{\!\!\!\!
\begin{array}{ll}
\{ 2x[i\!-\!1]\!+\!1, \, (y[i\!-\!1]\!-\!1)/2\}, &\mbox{if\ } -\!1\!\le\! x[i\!-\!1]\!<\!0 \\
 \{1\!-\!2x[i\!-\!1], \, (1\!-\!y[i\!-\!1])/2\}, &\mbox{if\ } 0\!\le \! x[i\!-\!1]\!\le \!1
\end{array}
\right.\label{eqn:baker_func0}
\end{align}

\begin{figure}[htb]
\centerline{
\includegraphics[width=2.4in]{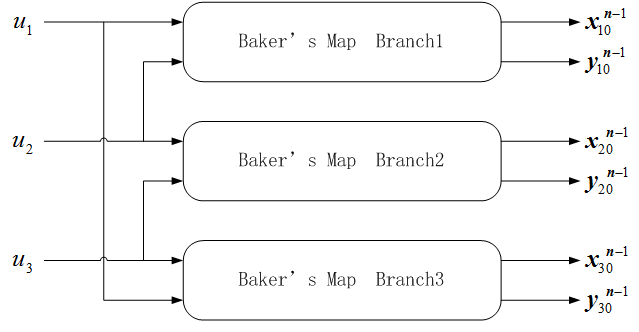}
}
\caption{The proposed triple-branch tail-biting analog codes.}
\label{fig:3branchBaker}
\end{figure}

Now consider building analog codes by engaging three baker's maps in a
looped tail-biting manner, as shown in Fig.
\ref{fig:3branchBaker}. A block of three real-valued symbols (e.g.
pixels in an image), $\{u_{1},u_{2},u_{3}\}$, is paired and fed into the three
branches as the initial states: $\{ u_{1},u_{2}\}$, $\{
u_{2},u_{3}\}$ and $\{ u_{3},u_{1}\}$. Each branch encoder
recursively performs the baker's map $F(\{x,y\})$ in
(\ref{eqn:baker_func0}), to generate additional $(n\!-\!1)$ pairs
of states (in addition to the initial states):
\begin{align}
 \{x_{j}[i],y_{j}[i]\}
=&F(\{x_{j}[i\!-\!1],y_{j}[i\!-\!1]\})\nonumber \\
=&F^{i}(\{x_{j}[0],y_{j}[0]\}), 
\label{eqn:baker_recursive}
\end{align}
where the subscript $j=1,2,3$ denotes the $j$th branch, and
$i=1,2,\cdots, n\!-\!1$ denotes the time index of the states.

The collection of all the states from $0$ to $(n\!-\!1)$,  
$\{\mathbf{x_{1}}{}^{n-1}_{0},
\mathbf{y_{1}}{}^{n-1}_{0}\}$, $\{\mathbf{x_{2}}{}^{n-1}_{0},
\mathbf{y_{2}}{}^{n-1}_{0}\}$ and $\{\mathbf{x_{3}}{}^{n-1}_{0},
\mathbf{y_{3}}{}^{n-1}_{0}\}$, forms the sub-codeword for the first, second, and third branch, respectively. Altogether $6n$ symbols are generated\footnote{The
$6n$ symbols include two copies of the source symbols $u_1,u_2,u_3$. It 
is also possible to transmit the systematic symbols only once,
which will lead to a codeword of $(6n\!-\!3)$ symbols and code rate
$1/(2n-1)$.}, corresponding to the three source symbols $\{
u_1,u_2, u_3\}$. Hence, the code is a $(6n, 3)$ systematic code
with a rate of $r=1/(2n)$.

\subsection{Transmission of Analog Symbols through CASK and CQAM}

The codewords are transmitted through the noisy channel. Each
analog symbol takes value from -1 to 1, and is 
modulated as variations in the amplitude of a carrier wave, in a
way similar to digital amplitude shift keying (ASK). 
The only difference is that an $m$-ary ASK only allows
a fixed set of  $m$ discrete amplitude values to be valid (e.g.
$-\Delta$, $-\Delta/3$, $\Delta/3$ and $\Delta$ for 4-ary ASK),
whereas in this {\it continuous ASK} (CASK), amplitudes may be any
real value between $-\Delta$ and $\Delta$. Hence,
our CASK may be regarded as an $\infty$-ary ASK.

From the communication theory, two $m$-array ASK modulations can
be packed to form a $m^2$-ary quadrature amplitude modulation
(QAM). When the two carrier waves use sinusoids that are out of
phase with each other by $90^o$ (termed $I$-channel and
$Q$-channel respectively), then the $m^2$-ary QAM achieves twice
the data rate ($2 log_2 m$ bits/symbol) on the same bandwidth as a
single $m$-ary ASK. Likewise, we can also pack two CASK to form a
{\it continuous QAM} (CQAM) to double our data rate. From the
perspective of signal space,  this is to take two of our
real-valued symbols to form a complex-valued symbol, and projected it
onto an $\infty$-ary QAM.

Mathematically, the noisy reception at the decoder is: 
\begin{equation}\label{eq:noisy cw}
\left\{\!\!\!
\begin{array}{ll}
I\mbox{-channel:}  & \mathbf{Rx_{j}}{}^{n-1}_{0} =  \mathbf{x_{j}}{}^{n-1}_{0}  +  \mathbf{n_{xj}}{}^{n-1}_{0} \\
Q\mbox{-channel:} & \mathbf{Ry_{j}}{}^{n-1}_{0} =  \mathbf{y_{j}}{}^{n-1}_{0}  +  \mathbf{n_{yj}}{}^{n-1}_{0} \\
\end{array}
\right. \ j\!=\!1,2,3,
\end{equation}
where $\mathbf{n_{xj}}{}_{0}^{n-1}$ and $\mathbf{n_{yj}}{}_{0}^{n-1}$ are
noise sequences. If we adopt the additive white Gaussian noise
(AWGN) channel model, then these noise samples follow independent
and identically distributed (i.i.d.) Gaussian distribution $\sim
{\cal N}(0, N_0/2)$.

\subsection{Decoder}

When two copies of the systematic symbols,
 $x_1[0]\!=\!y_3[0]\!=\!u_1$, $x_2[0]\!=\!y_1[0]\!=\!u_2$,
$x_3[0]\!=\!y_2[0]\!=\!u_3$, are transmitted, it is advisable to first perform
maximum ratio combining (MRC) before proceeding to the actual decoding.
On a homogeneous channel such as AWGN channel, MRC is equivalent
to equal gain combining (EGC):
\begin{align}
Rx'_{1}[0]= Ry'_{3}[0]=  \frac{Rx_{1}[0]+Ry_{3}[0]}{2},\\
Rx'_{2}[0]=Ry'_1[0]= \frac{ Rx_{2}[0]+Ry_{1}[0]}{2}, \\
Rx'_{3}[0]=Ry'_2[0]= \frac{ Rx_{3}[0]+Ry_{2}[0]}{2},
\end{align}
where the apostrophe $'$ denotes the symbols after MRC. For
convenience, we abuse the notation, and omit the apostrophe in
$Rx_j[0]$ and $Ry_j[0]$ in the following discussion.

The maximum-likelihood decoder tries to make the best decision of
the initial states, $u_1, u_2, u_0$, based on the noisy
observation of a sequence of states. From the definition of
baker's map in (\ref{eqn:baker_func0}), a later $x$-state  $x_j[i]$
can be deduced unequivocally from a previous one $x_j[i-1]$, but
not the other way around; and the same holds for the $y$-states.
The ambiguity in the backward derivation is caused by the unknown
sign of the previous $x$-state $x_i[i-1]$. Hence, to facilitate
decoding, we introduce a sign sequence $\mathbf{s_j}{}_0^{n-1}$ for
$\mathbf{x_j}{}_0^{n-1}$  (the signs of  $\mathbf{y_j}{}_0^{n-1}$ are
irrelevant):
\begin{align}
s_j[i]=sign(x_j[i]), \ \  i\!=\!0,1,...,n\!-\!1,\ \ j\!=\!1,2,3.
\end{align}

With the sign sequence established, we can 
establish a one-to-one mapping between $x_j[i]$ and $x_j[i\!-\!1]$ and between  $y_j[i]$ and $y_j[i\!-\!1]$ 
\begin{align}
\hspace{1cm}\left\{
\begin{array}{l}
x_j[i\!-\!1]= -\frac{1}{2}s_j[i\!-\!1](x_j[i]-1), \\
y_j[i\!-\!1]=-2s_j[i\!-\!1]y_j[i]+1.
\end{array}
\right.
\end{align}

Recall that the baker's map is a piece-wise linear function. With
each time evolution, the number of segments doubles, but linearity
preserves within each (new) segment.
Hence, one can rewrite the encoding function in
(\ref{eqn:baker_func0}) by directly establishing a linear relation
between the $i$th state $\{x_{j}[i],y_{j}[i]\}$ with the initial
state  $\{x_{j}[0],y_{j}[0]\}$ in each segment: 
\begin{align}
\left\{
\begin{array}{l}
x_{j}[i] = a_{j}[i]x_{j}[0]+b_{j}[i],\\
y_{j}[i] = c_{j}[i]y_{j}[0]+d_{j}[i],
\end{array}
\right. j\!=\!1,2,3.
\label{eqn:linear}
\end{align}
In general, the values of the parameters
$a_{j}[i],b_j[i],c_j[i],d_j[i]$ not only depend on the time index
$i$ but also on which segment $x_j[i]$ and $y_j[i]$ fall in.
Observe that the sign sequence  $\mathbf{s_j}{}_0^{n-1}$ actually
serves as the natural label for all the segments, namely,  $s_j[0]\in\{-1,+1\}$ specifies the two segments at
time index $i=1$, $(s_j[0]s_j[1])\in\{-1-1,-1+1,+1-1,+1+1\}$ specifies
the four segments at time index $i=2$, and so on.  Thus, carefully
arranging the sign sequence, we can derive the parameters
$a_{j}[i],b_j[i],c_j[i],d_j[i]$ in a unified recursive form across all the segments:
\begin{align}
\left\{
\begin{array}{l}
a_{j}[i]  =  -2s_{j}[i\!-\!1]a_{j}[i\!-\!1],\\
b_{j}[i]  =  1-2s_{j}[i\!-\!1]b_{j}[i\!-\!1],\\
c_{j}[i]  =  - 0.5\, s_{j}[i\!-\!1]c_{j}[i\!-\!1],\\
d_{j}[i] =  0.5\, ( s_{j}[i\!-\!1]-s_{j}d_{j}[i\!-\!1]),\\
\end{array}
\right. j\!=\!1,2,3.
\label{eqn:abcd}
\end{align}

Given the linear relation in (\ref{eqn:linear}) and (\ref{eqn:abcd}), an efficient ML decoding can be derived to obtain an estimation of the information bits $\{ \tilde{u}_{1}, \tilde{u}_{2},
\tilde{u}_{3}\}$.
\begin{align}
 & \{\tilde{u}_{1},\tilde{u}_{2},\tilde{u}_{3}\}\nonumber \\
=& \underset{-1\leq u_{1},u_{2},u_{3}\le 1}{\arg\max}\!\!
 \Pr\big(\mathbf{Rx_{1}}{}^{n-1}_{0},\mathbf{Ry_{1}}{}^{n-1}_{0},\mathbf{Rx_{2}}{}^{n-1}_{0}, \mathbf{Ry_{2}}{}^{n-1}_{0},\nonumber \\
&\ \ \ \ \ \ \ \ \ \ \ \ \ \ \ \ \ \ \ \ \mathbf{Rx_{3}}{}^{n-1}_{0}, \mathbf{Ry_{3}}{}^{n-1}_{0} \mid  u_{1},u_{2},u_{3}\big) \nonumber \\
= &  \!\underset{-1\leq u_{1},u_{2},u_{3}\leq
1}{\arg\max} \prod_{i=0}^{n-1}\! \Big\{\!\!
  \Pr\big(Rx_{1}[i]\mid u_1) \Pr\big(Ry_{1}[i] \mid u_2\big)\nonumber \\
&\ \ \ \ \ \  \ \ \ \ \ \ \ \cdot
  \Pr\big(Rx_{2}[i]\mid u_2) \Pr\big(Ry_{2}[i] \mid u_3\big) \nonumber \\
&\ \ \ \ \ \ \ \ \ \ \ \ \ \cdot \Pr\big(Rx_{3}[i]\mid u_3\big) \Pr\big(Ry_{3}[i] \mid u_1 \big)\Big\} \label{eqn:ML_1} \\
= &  \underset{-1\leq u_{1},u_{2},u_{3}\leq
1}{\arg\min}  \sum_{i=0}^{n-1}\Big\{
\big(Rx_{1}[i]\!-\!x_{1}[i]\big)^2 \!+\!
(Ry_{1}[i]\!-\!y_{1}[i]\big)^2 \nonumber \\
&\ \ \ \ \ \ \ \ \ \ \ \ + \big( Rx_{2}[i]\!-\!x_{2}[i]\big)^2 \!+\!
(Ry_{2}[i]\!-\!y_{2}[i]\big)^2
\nonumber \\
& \ \ \ \ \ \ \ \ \ \ \ \ + \big( Rx_{3}[i]\!-\!x_{3}[i]\big)^2 \!+\!
\big(Ry_{3}[i]\!-\!y_{3}[i]\big)^2  \Big\},
\label{eqn:ML}
\end{align}
where the equality in (\ref{eqn:ML_1}) is due to the independence
of the noise (i.e. memoryless channel), and the equality in
(\ref{eqn:ML}) is due to the Gaussianity of the noise. Using the
segmented linear function expressions in (\ref{eqn:linear}) and
noting that $x_1[0]=y_3[0]=u_1$, $x_2[0]=y_1[0]=u_2$, and
$x_3[0]=y_2[0]=u_3$,  we can further simplify (\ref{eqn:ML}) to: 
\begin{align}
&\{\tilde{u}_{1},\tilde{u}_{2},\tilde{u}_{3}\}
=\underset{-1\leq u_{1},u_{2},u_{3}\leq
1,\ \mathbf{s_{1}}{}_{0}^{n-2},\mathbf{s_{2}}{}_{0}^{n-2},\mathbf{s_{3}}{}_{0}^{n-2}}{\arg\min}
\sum_{i=0}^{n-1}\nonumber \\
& \Big\{
\big(Rx_{1}[i]\!-\!a_{1}[i]u_{1}\!-\!b_{1}[i]\big)^{2} +
\big(Ry_{1}[i]\!-\!c_{1}[i]u_{2}\!-\!d_{1}[i]\big)^{2}  \nonumber \\
& \big(Rx_{2}[i]\!-\!a_{2}[i]u_{2}\!-\!b_{2}[i]\big)^{2} +
\big(Ry_{1}[i]\!-\!c_{2}[i]u_{3}\!-\!d_{2}[i]\big)^{2}  \nonumber \\
& \big(Rx_{3}[i]\!-\!a_{3}[i]u_{3}\!-\!b_{3}[i]\big)^{2} +
\big(Ry_{3}[i]\!-\!c_{3}[i]u_{1}\!-\!d_{3}[i]\big)^{2}  \Big\}
\label{eqn:ML_2}
\end{align}
where the parameters $a_{j}[i],b_{j}[i], c_{j}[i],d_{j}[i]$ are
given in (\ref{eqn:abcd}).

The quadratic minimization problem in (\ref{eqn:ML_2}) can be solved
by taking the derivatives with respect to $u_{1}$, $u_{2}$ and
$u_{3}$, respectively. The ``global'' optimal solutions
$u_{1}^{\ast}$, $u_{2}^{\ast}$ and $u_{3}^{\ast}$ that minimize
(\ref{eqn:ML_2}) are:
\begin{align}
u_{1}^{\ast}\! =\!  \frac{\sum_{i=0}^{n-1} (Rx_{1}[i]a_{1}[i]\!+\!Ry_{3}[i]c_{3}[i]\!-\!a_{1}[i]b_{1}[i]\!-\!c_{3}[i]d_{3}[i])}{\sum_{i=0}^{n-1}(a_{1}^{2}[i]+c_{3}^{2}[i])}\nonumber\\
u_{2}^{\ast} \! = \! \frac{\sum_{i=0}^{n-1}(Rx_{2}[i]a_{2}[i]\!+\!Ry_{1}[i]c_{1}[i]\!-\!a_{2}[i]b_{2}[i]\!-\!c_{1}[i]d_{1}[i])}{\sum_{i=0}^{n-1}(a_{2}^{2}[i]+c_{1}^{2}[i])}\nonumber\\
u_{3}^{\ast} \!=\!
\frac{\sum_{i=0}^{n-1}(Rx_{3}[i]a_{3}[i]\!+\!Ry_{2}[i]c_{2}[i]\!-\!a_{3}[i]b_{3}[i]\!-\!c_{2}[i]d_{2}[i])}{\sum_{i=0}^{n-1}(a_{3}^{2}[i]+c_{2}^{2}[i])}
\label{eqn:MLresult1}
\end{align}

It should be noted that the ``global'' optimal decisions
$u^{\ast}_1, u^{\ast}_2, u^{\ast}_3$ in  (\ref{eqn:MLresult1}) are
not always the feasible solution, since they may fall outside the
support range, i.e. the respective linear segment of length 
$\frac{1}{2^{n-1}}$.
To account for the boundary conditions, note that, for $j=1,2,3$, 
\begin{equation}
\label{eqn:fn}
-1\le x_{j}[n-1]=a_{j}[n-1]u_{j}\!+\!b_{j}[n-1]\le 1, 
\end{equation}
which leads to
\begin{align}
& \min\left( \frac{-\!b_{j}[n\!-\!1]\!+\!1}{a_{j}[n\!-\!1]}, \ \frac{-\!b_{j}[n\!-\!1]\!-\!1}{a_{j}[n\!-\!1]}\right)
\le   u_j \nonumber \\
& \ \ \ \ \ \ \ \ \ \ \ \ \ \ \le  \max\left(\frac{-\!b_{j}[n\!-\!1]\!+\!1}{a_{j}[n\!-\!1]} , \frac{-b_{j}[n\!-\!1]\!-\!1}{a_{j}[n\!-\!1]}\right).
\label{eqn:boundary}
\end{align}
Combining the quadratic minimal solution in (\ref{eqn:MLresult1}) and
the boundary solution in (\ref{eqn:boundary}), the ML decoder will produce the following
decision:
\begin{align}
\tilde{u}_{j}\!= &
\left\{\!
\begin{array}{l}
\min\Big(\frac{-b_{j}[n-1]+1}{a_{j}[n-1]},  \frac{-b_{j}[n-1]-1}{a_{j}[n-1]}\Big), \\
\ \ \ \ \ \ \ \ \ \ \ \ \ \ \ \ \ \ \
\mbox{if\ } u_{j}^{\ast}\!\!<\!\!\min\Big(\frac{-b_{j}[n-1]+1}{a_{j}[n-1]},\frac{-b_{j}[n-1]-1}{a_{j}[n-1]}\Big), \\
\max\Big(\frac{-b_{j}[n-1]+1}{a_{j}[n-1]}, \frac{-b_{j}[n-1]-1}{a_{j}[n-1]}\Big),\\
\ \ \ \ \ \ \ \ \ \ \ \ \ \ \ \ \ \ \   \mbox{if\ }
u_{j}^{\ast}\!\!>\!\!\max\Big(\frac{-b_{j}[n-1]+1}{a_{j}[n-1]},\frac{-b_{j}[n-1]-1}{a_{j}[n-1]}\Big), \\
u_{j}^{\ast},  \ \ \ \ \ \ \ \ \ \ \ \ \ \  \ \mbox{otherwise},
\end{array}
\right.\nonumber\\
& \ \ \ \ \ \ \ \ \ \ \ \ \ \ \ \ \ \ \ \ \ \ \ \ \ \  \mbox{for\ }j=1,2,3.
\label{eqn:MLresult2}
\end{align}

%
%

\section{Image Transmission via Analog Codes}
\label{sec:simulation}

Analog codes are most useful for transmitting analog sources as shown in
Fig. \ref{fig:analogDigital}, but 
they can also be used to transmit digital data, and
especially digitized images.

To illustrate, consider a monochrome image, the $256\!\times\!256$ (pixel) Lena, where each pixel is represented by a byte valued between 0 and
255. In the conventional digital transmission paradigm, all the
bits are assembled into a bit-stream ($65536\!\times\! 8$ bits altogether), and digitally coded and
modulated. In the proposed analog transmission paradigm, the 
pixels are viewed as stream of real-valued analog symbols ($65536$ symbols in the range of $[-1,1]$, i.e. each pixel $0\le x\le 255$ is linearly scaled to $[-1,1]$ via $x\to \frac{x\!-\!128}{128}$), and then coded
by an analog code. We simulate and compare the analog system with 
some of the best-known digital systems:

1). The analog system consists of the proposed $(12,3)$ 3-branch
tail-biting baker's map code with code rate 1/4. The encoded
symbols are transmitted via the continuous ASK modulation ($65536\!\times\! 4$ total symbols). 

2). The digital system consists of a digital turbo code, one of
the best error correction codes known to date, and the ASK modulation. We
consider $(4096,2048)$ 16-state turbo codes with constituent
convolutional codes  $(1,\frac{1+D+D^2+D^3}{1+D+D^3})$ and code
rate 1/2, and 16-ary ASK. Thus, with the default 8-bit quantization per pixel, the overall bandwidth expansion is $\frac{1}{8}\frac{1}{2}\log_216\!=\!1/4$, which is the same as the analog system (i.e. $65536\!\times\! 8\!\times\! 2/4=65536\!\times\!4$ symbols). At the decoder, the 16-ary ASK is softly demodulated, and the resultant log-likelihood ration (LLR) is passed to the soft iterative turbo decoder which uses the BCJR algorithm as the sub-decoders. Six iterations are performed before the decoder outputs hard decisions. 

The transmission quality is evaluated using  the  mean square
error (MSE) between the original $m\times l$ monochrome image $I$
and the reconstructed image $K$,
\begin{align}
\mbox{MSE}\!=\frac{1}{ml} \sum_{i=0}^{m-1}\sum_{j=0}^{l-1} (I_{i,j} - K_{i,j})^2,
\end{align}
 as well as the peak signal-to-noise ratio (PSNR),
\begin{align}
\mbox{PSNR}=20\log_{10} (I_{max}/\sqrt{\mbox{MSE}})\ \ \ (dB),
\end{align}
where $I_{max}$ is the maximum possible pixel value of the image
(e.g. 255 for 8-bit quantized monochrome images).

\begin{figure}[htb] 
\vspace{-0.2cm}
\centerline{
\includegraphics[width=3in,height=2in]{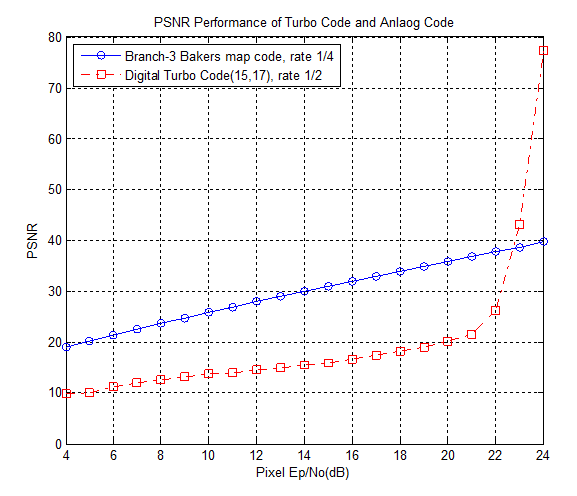}
\vspace{-0.2cm}
}
\caption{PSNR (dB) of the Lena Picture transmitted by the analog system and the digital system (digital turbo codes)}
\label{fig:psnr}
 \vspace{-0.2cm}
\end{figure}

Fig. \ref{fig:psnr} plots the PSNR performance (dB) of the digital system and the analog system on AWGN channels with signal-to-noise ratio (SNR) measured in terms of $E_p/N_0$ (dB), where $E_p$ is the average energy per pixel in the original image. 
Because of the bandwidth expansion in digital system (1 pixel becomes 8 bits), the digital turbo code cannot get to its waterfall region until after a rather high $E_p/N_0$ of $22$ dB. As a consequence,    
the proposed analog code noticeably outperforms the digital turbo code for a wide range of channel conditions. In general, a PSNR of 30 dB or more is reckoned as good quality image. This is achieved by the analog system at $E_p/N_0=14$ dB, but is not achieved by the digital system until  $E_p/N_0>22$ dB!

\begin{figure}[htb]\label{fig:lena15db}
\begin{center}
\includegraphics[totalheight=1.6in]{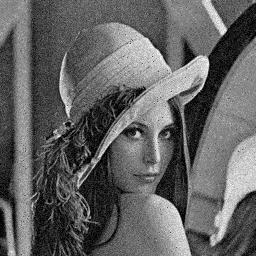}
\hspace{0.2cm}
\includegraphics[totalheight=1.6in]{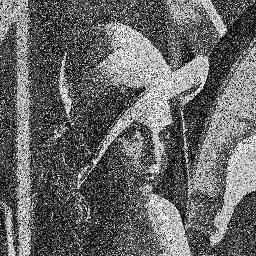}
\\
\vspace{0.2cm}
\includegraphics[totalheight=1.6in]{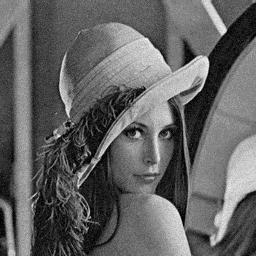}
\hspace{0.2cm}
\includegraphics[totalheight=1.6in]{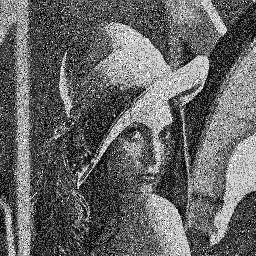}
\\
\vspace{0.2cm}
\includegraphics[totalheight=1.6in]{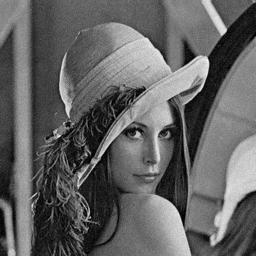}
\hspace{0.2cm}
\includegraphics[totalheight=1.6in]{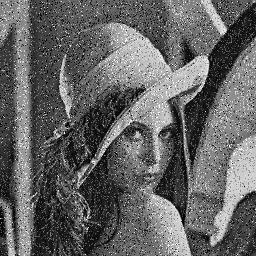}
\\
\vspace{0.2cm}
\includegraphics[totalheight=1.6in]{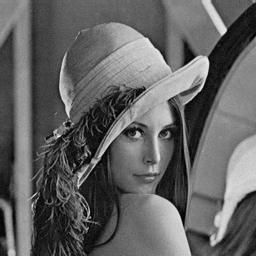}
\hspace{0.2cm}
\includegraphics[totalheight=1.6in]{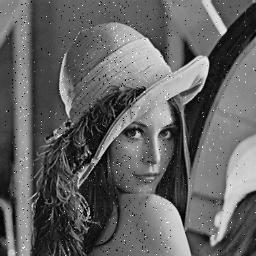}
\\
\vspace{0.2cm}
\includegraphics[totalheight=1.6in]{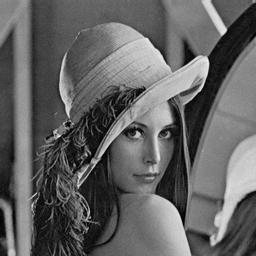}
\hspace{0.2cm}
\includegraphics[totalheight=1.6in]{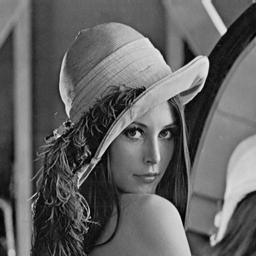}
\vspace{-0.5cm}
\end{center}
\caption{Comparison of recovered $256\!\times\! 256$ (pixel) Lena Picture by the proposed analog code (left) and the digital turbo code (right), for $E_p/N_o$ at $10$, $14$, $18$, $22$ and $24$ dB (from top to bottom). $E_p$ stands for energy per pixel. Both systems have a bandwidth expansion of $1:4$.}
\label{fig:lena}
\end{figure}

To provide a visual feel of the transmission quality,  Fig. \ref{fig:lena} further demonstrates the reconstructed
images for the two systems at $E_p/N_0$ of 10, 14, 18, 22 and 24 dB, respectively. We see that the digital system (right column) still experiences annoying pepper-and-slat errors at a high $E_p/N_0$ of 22 dB, whereas the analog system (left column) can deliver quality image for as low as 14 dB.  

The advantages of the analog system are rather obvious, including the capability of delivering good quality on  poor channel conditions, graceful performance degradation, and considerably lower complexity. The last is particularly noteworthy. In the digital system, soft-demodulation and soft turbo decoding  are both very complex and time-consuming, and involve nonlinear operations (e.g. the $max^*$ operation in the BCJR decoding). In comparison, the proposed analog code entails only a few simple linear operations. Further, the digital turbo code has a considerably longer block length (2048 bits or 256 pixels) than the analog system (3 pixels), and hence requires considerably longer memory consumption and delay. 

Our simulations are currently run over raw or TIFF (tagged image file format) images, a  popular lossless format especially for high color-depth images. Although many consumer electronics use lossy JPEG format, raw images which allow editing and re-saving without losing image quality play an important role in image archive and especially in bio-medical applications. Further, the proposed analog codes can also be extended to compressed images, as there is no fundamental conflict between analog error correction coding and compression. For example, in a JPEG image, the DCT (discrete cosine transform) coefficients (which are by nature real-valued) are each represented as a binary bit stream in the digital coding systems; but they can be directly taken in as analog symbols in the analog coding systems. Analog coding can also provide data accuracy as required; for example, if an analog code operating on sources bounded between $[-1,1]$ can provide an MSE $\Delta(<1)$, then it can on average guarantee the accuracy of $|\frac{1}{2}\log_2\Delta|$ binary bits after the decimal. Hence, with careful arrangement, analog coding schemes can also find good use in transmitting compressed images (including the control data and the efficients). 
    



%
%

\section{Conclusion}
\label{sec:conclusion}

We have developed an efficient triple-branch tail-biting baker's map analog code. Using a simple linear-operation-based maximum likelihood decoding scheme, we apply the code to image transmission, and show that, despite its considerably lower complexity, memory consumption and delay, the analog code actually significantly outperforms turbo-code-based digital systems! We conclude by promoting analog coding as a new and potentially very rewarding technology for transmitting images as well as other analog sources. 



\begin{thebibliography}{99}
\small  

\bibitem{bib:analogTransmit}
C. W. Helstrom, ``Topics in the transmission of continuous
information,'' Westinghouse Res. Lab., Res. Rep. 64-8C3-522-Rl. Aug. 21, 1964.

\bibitem{bib:Mars81}
T. G. Marshall, Jr.. ``Real number transform and convolutional
codes,'' {\it Proc. 24th Midwest Symp. Circuits Sys.}, 
Editor: S. Kame, Albuquerque, NM, June 29-30, 1981

\bibitem{bib:Wolf83a}
J. K. Wolf, ``Analog codes,'' {\it IEEE Intl. Conf. Comm,} Boston, MA, USA, 
June, 1983, pp. 310-312.


\bibitem{bib:Wolf83b}
J. K. Wolf, ``Redundancy, the discrete Fourier transform, and
impulse noise cancellation,'' \emph{IEEE Trans. Comm.}, Vol.
COM-31, No. 3, pp. 458-461, March 1983

\bibitem{bib:chaos}
S. Wiggins, {\it Introduction to Applied Nonlinear Dynamical Systems and Chaos,}
 Springer Publisher, 1993.

\bibitem{bib:vardy}
N. Anthia, A. Vardy, ``Analog codes on graphs,'' submitted to {\it
IEEE Trans. Inform. Theory}, arXiv:cs/060808vc1.


\bibitem{bib:DCT}
J.-L. Wu and J. Shiu, ``Discrete cosine transform in error control
coding'', {\it IEEE Trans. Comm.}, pp. 1857-1861, May 1995.

\bibitem{bib:tentmapcode}
B. Chen and G. W. Wornell, ``Analog error-correcting codes based on
chaotic dynamical systems,'' \emph{IEEE Trans. Comm.}, vol 46,
Issue 7, July 1998, pp: 881-890

\bibitem{bib:cat}
K. Xie, P. Tan, B. C. Ng. and J. Li (Tiffany), ``Analog turbo
codes: A chaotic construction,'' \emph{IEEE Intl. Symp. Inf.
Theory}, 2009.

\bibitem{bib:mirror}
K. Xie, J. Li, ``Chaotic Analog Error Correction Codes: The 
Mirrored Baker's Codes,'' \emph{IEEE Global Telecom. Conf.}, 2010


\bibitem{bib:linearAnalog}
K. Xie and J. Li (Tiffany), ``Linear analog codes: The good and the bad,'' 
http://arxiv.org/abs/1105.1520 

\bibitem{bib:chua}
R. N. Mada (Guest editor), {\it J. Circuits Syst. Comput.- Special Isse on Chua's Circuit: A Paradigm for Chaos}, vol. 3, Mar. 1993 (Part I) and June 1993 (Part II). 

\end{thebibliography}
\end{document}